\documentclass[twocolumn,twoside,slac_two,floatfix]{revtex4}
\usepackage{graphicx}
\usepackage{fancyhdr}
\begin{document}

\title{First MINOS Results from the NuMI Beam}


\author{Nathaniel Tagg, for the MINOS Collaboration}
\affiliation{Tufts University, 4 Colby Street, Medford, MA, USA 02155}

\begin{abstract}
  As of December 2005, the MINOS long-baseline neutrino oscillation
  experiment collected data with an exposure of $0.93\times10^{20}$ protons on
  target.  Preliminary analysis of these data reveals a result
  inconsistent with a no-oscillation hypothesis at level of 5.8 sigma.
  The data are consistent with neutrino oscillations reported by
  Super-Kamiokande and K2K, with best fit parameters of $\Delta
  m^2_{23} = 3.05^{+0.60}_{-0.55}\times 10^-3$ and $\sin^22\theta_{23} =
  0.88^{+0.12}_{-0.15}$.
\end{abstract}

\maketitle

\thispagestyle{fancy}


\section{Introduction}
The MINOS long-baseline neutrino oscillation experiment
\cite{Thomson:2005hm} was designed to accurately measure neutrino
oscillation parameters by looking for $\nu_\mu$ disappearance. MINOS
will improve the measurements of $\Delta m^223$ first performed by the
Super-Kamiokande \cite{Fukuda:1998ah,Ashie:2005ik} and K2K experiments
\cite{Aliu:2004sq}.  In addition, MINOS is capable of searching for
sub-dominant $\nu_\mu \rightarrow \nu_e$ oscillations, can look for
CPT-violating modes by comparing $\nu_\mu$ to $\bar{\nu_\mu}$
oscillations, and is used to observe atmospheric neutrinos
\cite{Adamson:2005qc}.

The MINOS experiment uses a beam of $\nu_\mu$ created at Fermilab
National Laboratory and directed at the Soudan mine in northern
Minnesota, a distance of 735~km. The composition and energy spectrum
of the beam is measured in two detectors, the Near (1~km downstream
from the target) and the Far (735~km downstream), allowing for
precision measurements of the spectral distortion of the beam.

\section{The NuMI Neutrino Beam}
The Fermilab Main Injector has a minimum cycle time of 1.87~s, with a
maximum intensity of $4 \times 10^{13}$ protons per pulse for a
maximum average power of 0.4~MW of power on the target.  Protons are
extracted from the Main Injector in a single turn, taking $\sim 10 \mu$s.

The NuMI neutrino beam \cite{Kopp:2005bt} is created by directing these
protons onto a water-cooled segmented graphite
target.  Secondary pions from the proton-carbon interactions are
deflected into the forward direction by two parabolic focusing elements
(horns) before being allowed to decay in a 675~m evacuated decay
volume.  Undecayed secondaries are stopped by an absorber wall at the
end of this volume. The primary proton beam is monitored for position
and intensity on target.  The secondary hadrons and muons are monitored
for position and intensity to ensure good alignment and composition of
the beam.

The NuMI beam has been in operation since late 2004. Near the end of
2005, a total exposure of $0.93\times10^{20}$ protons were delivered
to the target. This exposure was used in the following analysis. At the end of the run period in March 2006, the maximum
intensity delivered to the target was in excess of $25 \times 10^{12}$
protons per pulse, with a maximum target power of 250~kW.

\section{The MINOS Detectors}

The MINOS Near and Far detectors are constructed to have nearly
identical composition and cross-section.  The detectors consist of
sandwiches of 2.54~cm thick steel and 1~cm thick plastic
scintillator, hung vertically.  The polyethylene scintillator is in
the form of 4~cm-wide strips, with a co-extruded TiO$_2$ reflective
coating. Along one side of each strip a groove holds a glued
wavelength-shifting optical fibre.  Scintillation light created in the
scintillator is caught by the fibre, shifted, and transported
efficiently to the end of the strip.  Clear readout fibres carry the
light to multi-anode photomultipliers for readout. 

The detectors act as tracking, sampling calorimeters. Strips in
adjacent planes are oriented orthogonally, allowing events to be
reconstructed in two transverse views.  Both detectors are equipped
with magnet coils which generate $\sim$1.2~T toroidal magnetic
fields, which act to contain long muon tracks and provide curvature
information for estimating energies.

The Near detector has a total mass of one kiloton, and 282 of these steel
planes, 153 of which are instrumented.  It uses sampling electronics
to distinguish neutrino events in time, due to the large instantaneous
intensity during a beam spill.  The Far detector masses 5.4 kilotons,
and consists of 485 steel planes, 484 of which are instrumented.
Because of the low rate in the Far detector, the strips are read out
via an 8-fold optical multiplexing.

\section{Data Selection}

Charged-current $\nu_\mu$ events in the MINOS detectors appear as long
muon tracks accompanied by short hadronic showers near the event
vertex.  The two other distinguishable event classes are neutral-current
events, which appear only as short, sparse hadronic showers, and
$\nu_e$ charged-current events, which appear as short, dense
electromagnetic showers. For this analysis, only charged-current
$\nu_\mu$ events are considered.

\begin{figure}[htb]
\centering
\includegraphics[width=80mm]{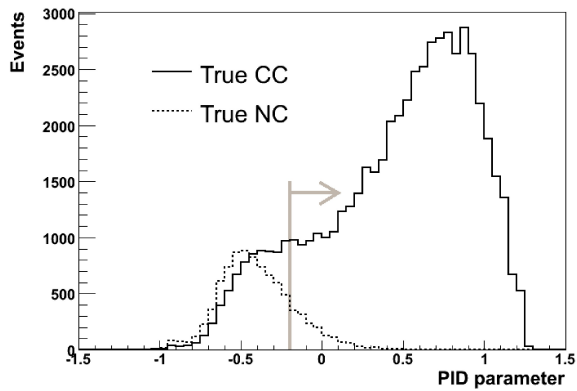}
\caption{The Particle ID Cut.  The solid line shows the distribution
  of Monte Carlo charged-current events and the dashed line shows the
  distribution of Monte Carlo neutral-current events as a function of
  the particle ID parameter used for the selection. The cut value is shown.}
 \label{f:cc_nc_cut}
\end{figure}

\begin{figure}[htb]
\centering
\includegraphics[width=80mm]{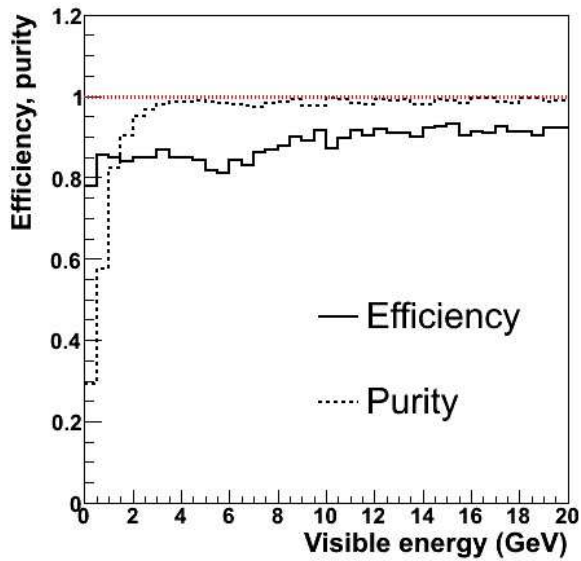}
\caption{Purity and Efficiency of the CC $\nu_\mu$ Event
  Selection. Efficiency and purity are shown as a function of
  reconstructed event energy, using Monte Carlo. Contamination events
  are from NC events, almost entirely at low reconstructed energy.}
 \label{f:eff_pur}
\end{figure}

Neutrino events are selected by first taking time-coincidence with the
beam spill. This is performed by hardware trigger in the Near
detector. In the Far detector candidate events are required to occur
at $2449 \pm 50\mu s$ after beam spill (the time-of-flight for
neutrinos traveling to Soudan).  Charged-current $\nu_\mu$ events
require a well-reconstructed track in both scintillator views, and
confinement of the track vertex to the defined fiducial volumes of
each detector.  The curvature of the track is selected to be
consistent with a $\mu^-$.  Finally, charged-current
$\nu_\mu$ events must pass a particle identification cut, which relies
on event length, the proportion of calorimetric energy in the shower,
and the track $dE/dx$.  These variables are assembled into a
probability density functions and used to create a likelihood-based
particle ID parameter shown in Figure \ref{f:cc_nc_cut}.  A cut is
performed on this variable to select a pure $\nu_\mu$ CC sample. The
total efficiency and purity of this selection is shown in Figure
\ref{f:eff_pur}.

For selected events, the neutrino energy is reconstructed by taking
the sum of the muon energy and the shower energy. Muon energy is found
using the range of the muon (if contained) with an accuracy of 6\%, or
by using the muon curvature (if uncontained) with an accuracy of 10\%.
The shower energy is found using calorimetry, with an uncertainty of
approximately $55\%/\sqrt{E}$.

\section{The Near Detector Spectrum}

The Near detector neutrino sample is used as the control in the
experiment, providing the unoscillated beam spectrum.
One of the advantages of the NuMI beam is that the target may be moved
relative to the horns, allowing a varying focus of the pions and
therefore different neutrino energies.  Three such settings are used
to create 'high', 'medium' and 'low' energy beams.  Although the bulk
of data was taken in the low energy configuration (to maximize
sensitivity at the oscillation minimum) data was also taken in the
other configurations to test beam modeling in the Near detector.

\begin{figure}[htb]
\centering
\includegraphics[width=80mm]{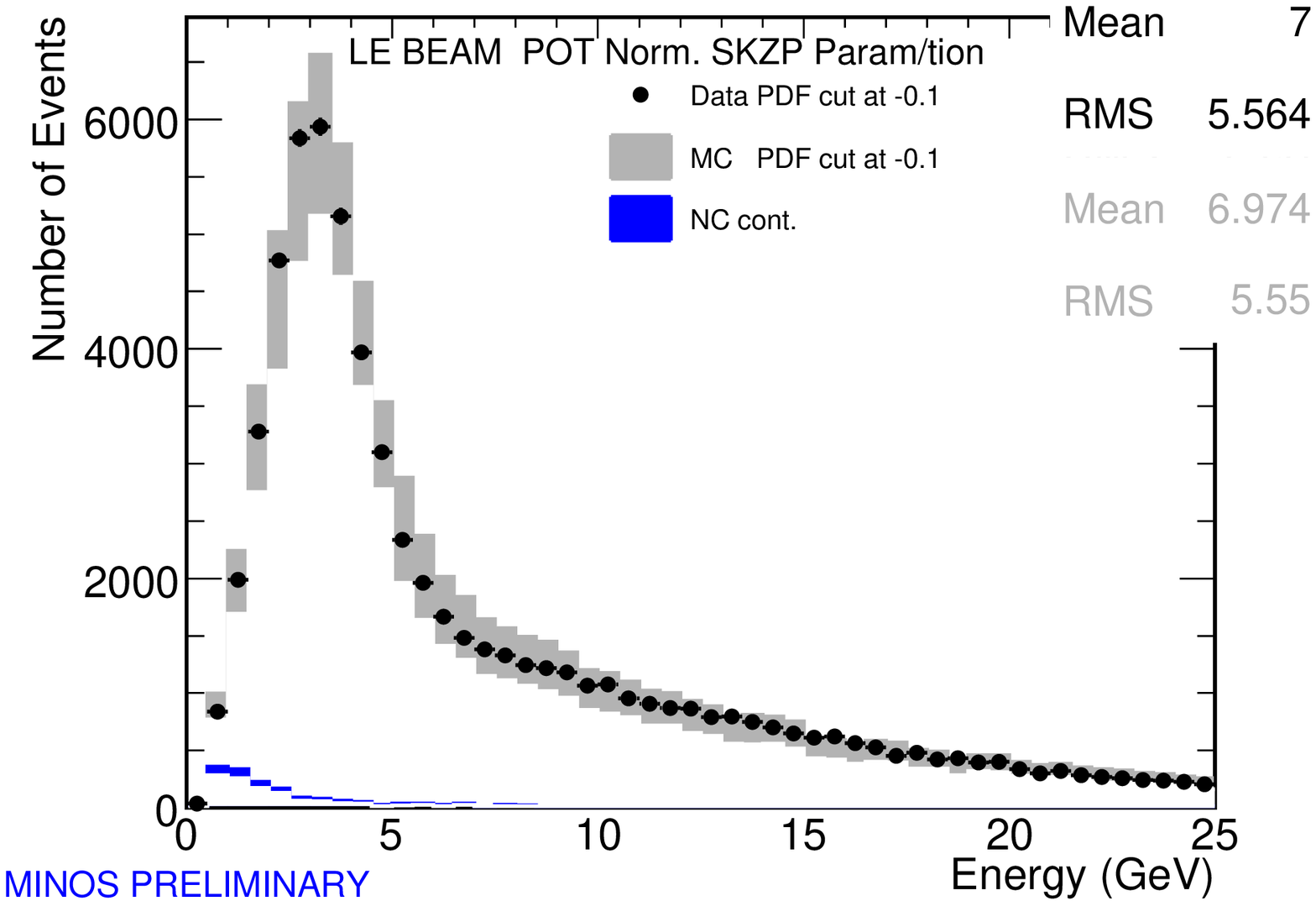}
\includegraphics[width=80mm]{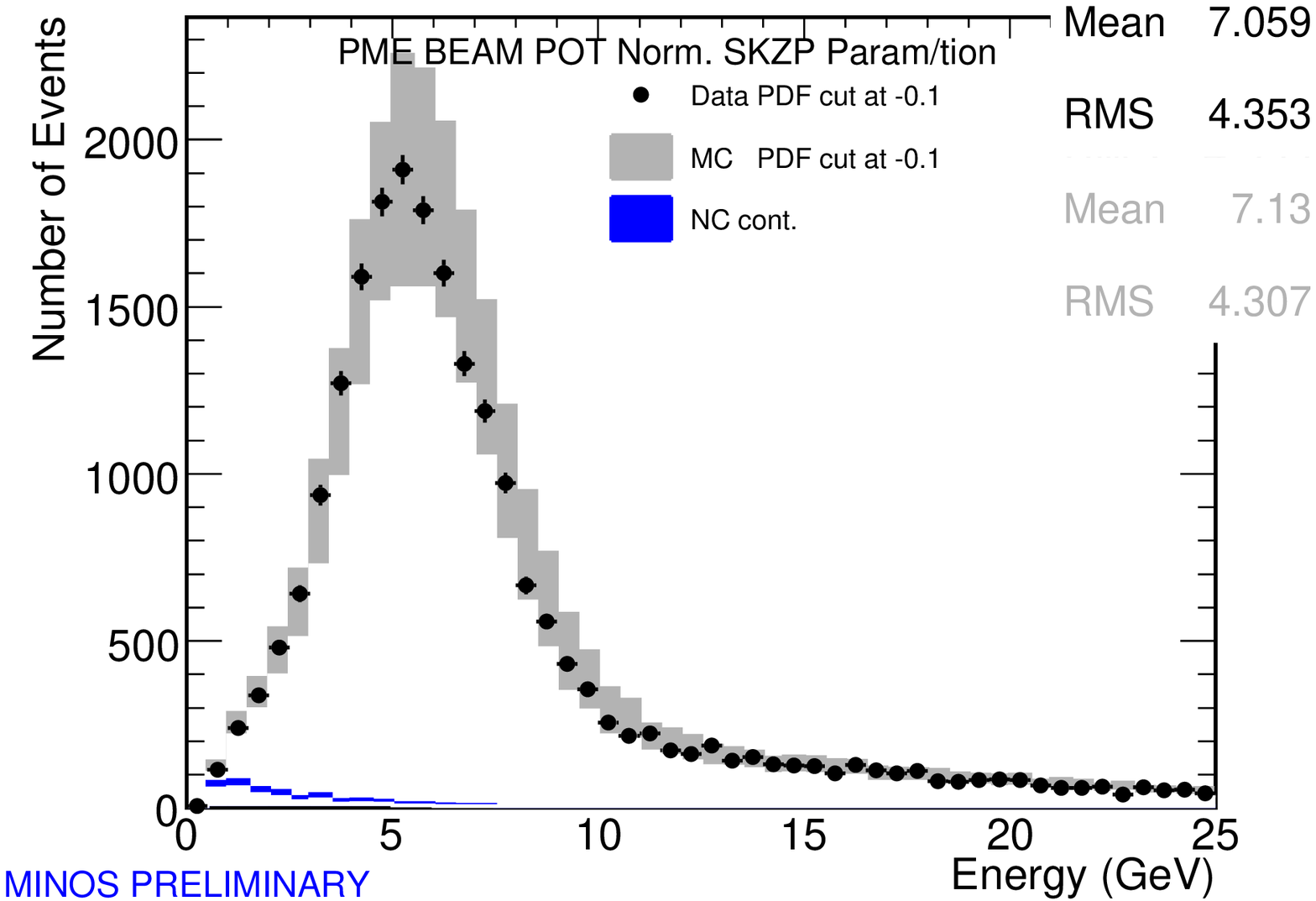}
\includegraphics[width=80mm]{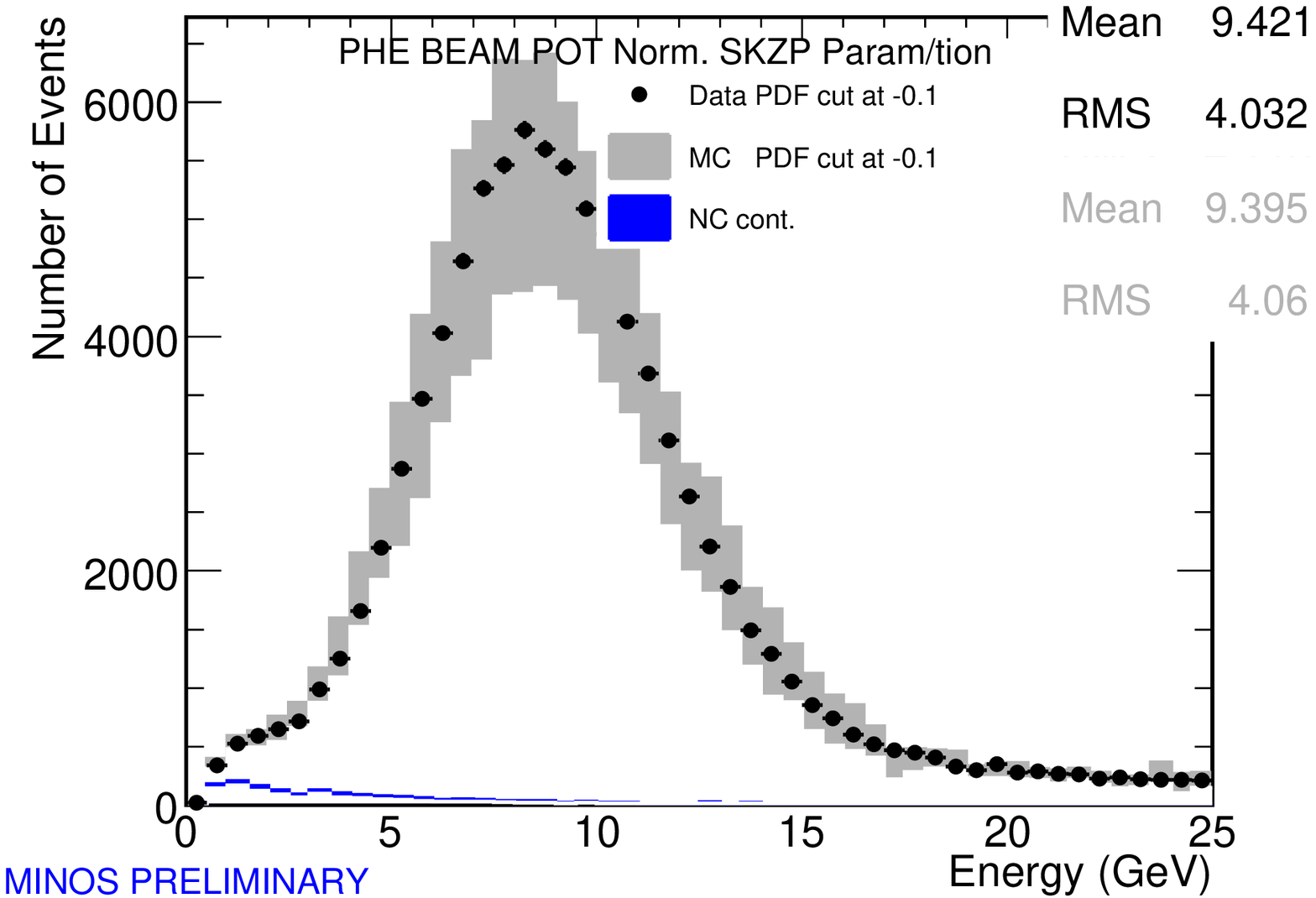}
\caption{Near Detector Neutrino Energy Spectra. The spectrum of
  observed charged current %%$\nu_\mu$ events is shown with the black
  dots. Hatched areas indicate the Monte Carlo prediction of the
  spectrum, after tuning. The predicted neutral current contamination
  is shown in blue. The top plot shows the data from the low energy
  beam configuration, the middle plot shows the medium energy
  configuration, and the lower plot shows the high energy
  configuration.}
 \label{f:nd_spec}
\end{figure}

Monte Carlo simulation of the beam predicted the neutrino flux for all
beam configurations to within 10\%.  The residual disagreement was a
strong function of beam configuration rather than neutrino energy,
implying that the simulation error was due to the modeling of the
secondary pions.  An empirical function was used to modify the
secondary pion simulation as a function of $x_f$ and $p_t$. This
function was fit to data from all configurations. These data and their
associated Monte Carlo spectra are shown in Figure \ref{f:nd_spec}.

The measured Near detector spectrum was used to predict the Far
detector spectrum. First, the observed Near detector reconstructed
energy spectrum was corrected for purity and efficiency to obtain the
true neutrino spectrum.

Second, the spectrum was corrected for the difference in solid angles
subtended by the Near and Far detectors.  The smaller range of angles
observed by the Far detector results in a narrower spectrum relative to
the Near detector. This difference can be robustly estimated by using
knowledge of the pion two-body decay kinematics and the geometry of
the NuMI beamline. Using the beam Monte Carlo, a matrix is created
that transforms a true Near detector spectrum into a true Far detector
spectrum. 

Last, the predicted true Far detector spectrum is simulated by Monte
Carlo to create an expected reconstructed neutrino spectrum which can
be compared to data.

\section{Far Detector Results}

\begin{figure}[htb]
\centering
\includegraphics[width=80mm]{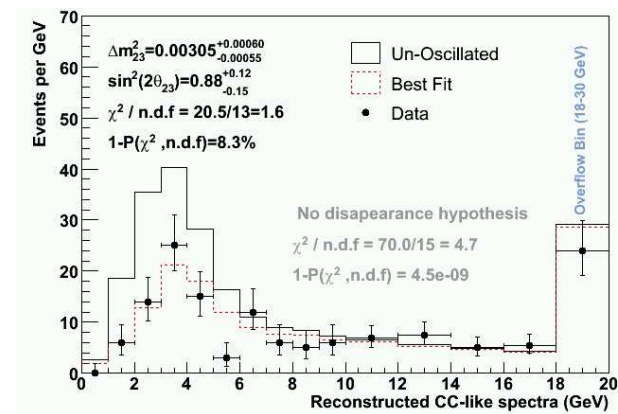}
\caption{{\it MINOS PRELIMINARY} Far Detector Energy Spectrum. The
  predicted spectrum in absence of oscillations is shown by the solid
  curve.  The dashed curve represents a fit to the data by an
  oscillation hypothesis with a best fit of $\Delta
  m^2_{23} = 3.05^{+0.60}_{-0.55}\times 10^-3$ and $\sin^22\theta_{23} =
  0.88^{+0.12}_{-0.15}$ }.
 \label{f:data}
\end{figure}

\begin{figure}[htb]
\centering
\includegraphics[width=80mm]{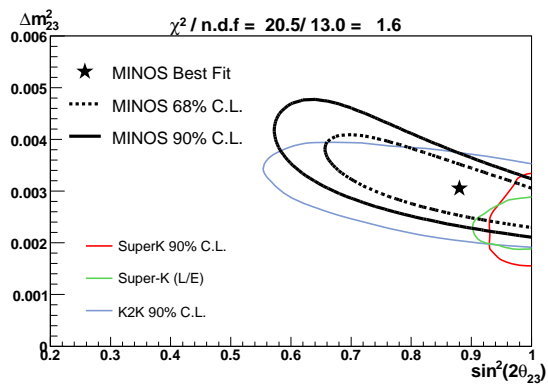}
\caption{{\it MINOS PRELIMINARY} Allowed Region. The black solid (dashed) line shows the MINOS
90\% (68\%) confidence limit in $\Delta m_{23}^2$ and
$\sin^22\theta_{23}$. The 90\% confidence regions from the
Super-K zenith-angle analysis \cite{Fukuda:1998ah}, the Super-K L/E
analysis \cite{Ashie:2005ik} and the K2K result \cite{Aliu:2004sq} are
shown in red, green, and blue respectively. The contour is described
by statistical errors only.}
 \label{f:contour}
\end{figure}

The observed reconstructed neutrino spectrum at the Far detector is
shown in Figure \ref{f:data}.  The data do not agree with the
unoscillated prediction; this null hypothesis is excluded at 5.8
sigma.  When fitted to a two-neutrino oscillation hypothesis, a fit is
found with reasonable probability within a parameter space shown in
Figure \ref{f:contour}.  In this fitting procedure, only statistical
errors were considered.

Systematic errors were estimated by taking Monte Carlo data samples
simulated near the best fit point for the data.  Systematic errors
were introduced to the Monte Carlo and not corrected in analysis. The
resulting shift in the best fit point is taken as the systematic error
on that parameter.  The systematic errors considered are shown in
Table \ref{t:sys}. These systematics are briefly discussed below.
\begin{description}
\item[Normalization] changes the overall expected number of events at
  the Far detector, based upon a combination of uncertainty on
  protons-on-target and the fiducial mass. This is known to 4\%.
\item[Muon energy scale] is the uncertainty on the average energy of
  muons due to incorrect steel density or magnetic field
  modeling. This is known to 2\%.
\item[Relative shower energy scale] is the degree of confidence to
  which the hadronic calorimetry in the Near detector matches that in
  Far.  The energy scale between these two detectors after calibration
  should agree to within 3\%.
\item[NC contamination] is the uncertainty on the number of neutral
  current background events erroneously accepted into the charged
  current sample. A conservative error of 30\% is used.
\item[CC Cross section] is allowed to vary in several ways: the KNO
  parameters governing the relative fraction of resonance to deep
  inelastic scattering fraction were allowed to change by 20\%, and the
  axial mass for quasi-elastic events and resonance events were
  allowed to change by 5\%. 
\item[Beam uncertainty] is the difference in measured parameters given
  that the data is fit without re-tuning the Monte Carlo beam
  simulation to the Near detector data. That is, an incorrect beam
  matrix was used to extrapolate the beam to find the shift in fit values.
\item[Intra-nuclear rescattering] gives the uncertainty due to poor
  modeling of the energy loss of pions scattered or re-absorbed as
  they leave the target nucleus. This energy cannot be detected, and
  so an uncertainty on this interaction creates an uncertainty on the
  absolute hadronic shower energy scale at both detectors.
\end{description}

\begin{table}
\centering
\begin{tabular}{lrr}
\hline
Description &  $\Delta \left(\Delta m^2_{23}\right)$ & $\Delta\left(\sin^22\theta_{23}\right)$\\
\hline
Normalization $\pm$ 4\%               & $0.63 \times 10^{-4}$ & 0.025 \\
Muon energy scale $\pm$ 2\%           & $0.14 \times 10^{-4}$ & 0.020 \\
Relative Shower energy scale $\pm3\%$& $0.27 \times 10^{-4}$ & 0.020 \\
NC contamination $\pm$ 30\%           & $0.77 \times 10^{-4}$ & 0.035 \\
CC cross-section uncertainties        & $0.50 \times 10^{-4}$ & 0.016 \\
Beam uncertainty                      & $0.13 \times 10^{-4}$ & 0.012 \\
Intra-nuclear re-scattering            & $0.27 \times 10^{-4}$ & 0.030 \\
\hline
Total Systematic Error  & $1.19 \times 10^{-4}$ & 0.063 \\
\hline
Total Statistical Error               & $6.4 \times 10^{-4}$ & 0.15 \\
\hline
\end{tabular}
\caption{{\it MINOS PRELIMINARY} Systematic Uncertainties. Systematic
  effect on the data are shown together with the shift incurred by not
  correcting for these systematics on the best-fit point of $\Delta
  m^2_{23}$ and $\sin^22\theta_{23}$.} 
\label{t:sys}
\end{table}

Table \ref{t:sys} also shows the sum of these systematic errors taken
in quadrature, as compared to the statistical errors.  It can be seen
that for this limited data set, MINOS is statistically limited.
Systematic errors have a negligible effect on the contour at this sensitivity.

\section{Conclusions}

MINOS has completed a preliminary analysis of the first $\sim 10^{20}$
protons on target, and found a result incompatible with no
oscillations at a significance level of 5.8 sigma.  The signal is
consistent with oscillations seen by Super-Kamiokande and K2K at the
90\% confidence limit.  Systematic errors are believed to be under
control at the limit of the current statistics.  This result will
significantly improve the world average on $\Delta m^2_23$.  

MINOS currently has a data sample of approximately $1.5 \times
10^{20}$ protons on target, and will be analyzing this expanded data
set for summer 2006.  MINOS intends to continue collecting data until
at least 2010, at which time it's physics sensitivity is expected to
be considerably improved.

\bigskip % extra skip inserted
\begin{acknowledgments}

  This work was supported by the U.S. Department of Energy, the
  U.K. Particle Physics and Astronomy Research Council, the
  U.S. National Science Foundation, the State and University of
  Minnesota, the Office of Special Accounts for Research Grants of
  the University of Athens, Greece, and FAPESP (Fundacao de Amparo a
  Pesquisa do Estado de Sao Paulo) and CNPq (Conselho Nacional de
  Desenvolvimento Cientifico e Tecnologico) in Brazil. We gratefully
  acknowledge the Minnesota Department of Natural Resources for
  their assistance and for allowing us access to the facilities of the
  Soudan Underground Mine State Park. 

  We particularly thank the Fermilab Beams Division and the rest of
  the Fermi National Laboratory for their work and support in
  creating, operating, and improving the NuMI beam. 
\end{acknowledgments}

\bigskip % extra skip inserted

\end{document}